\documentstyle[NATO,epsf,epsfig,namedreferences]{CRCKAPB} 

\begin{opening}
\title{Towards asteroseismology of the non-radial pulsating 
       sdB star PG\,1605$+$072}
\author{S. FALTER, U. HEBER}
\institute{Dr. Remeis-Sternwarte, Universit\"at Erlangen-N\"urnberg,\\ 
	   Sternwartstr. 7, D--96049 Bamberg, Germany}
\author{S. DREIZLER, S. L. SCHUH}
\institute{Institut f\"ur Astronomie und Astrophysik, Universit\"at T\"ubingen,\\
	   Sand 1, D--72076 T\"ubingen, Germany}
\author{O. CORDES}
\institute{Sternwarte der Universit\"at Bonn,\\
           Auf dem H\"ugel 71, D-53121 Bonn, Germany}
\end{opening}
\begin{document}
The recently discovered new class of sdB pulsators (sdBV) offers a powerful 
possibility for the investigation of their interior and thus their evolutionary 
history. The first step towards applying asteroseismologic tools is the 
identification of pulsation modes. The best target for such an investigation is 
PG\,1605$+$072 because of 
its outstanding properties among the sdBVs: it is the brightest object (B = 13
mag), it has the longest periods ($\sim$ 500 s) and the largest variations (0.2
mag in the optical) as well as the richest pulsation spectrum with more than 50
modes (see Kilkenny {\it et al.} 1999). The rather rapid rotation 
(${\rm v}\sin{i}=39\,{km}/{s}$, Heber {\it et al.} 1999) complicates the 
interpretation of the pulsation spectrum due to non-linear mode splitting.

We want to accomplish the mode identification by measuring both the light and
the radial velocity curve for PG\,1605$+$072. The analysis of equivalent width 
changes of the Balmer lines and of line profile variations (as a future task in
preparation) will support the aim of the project. Moreover, we want to determine
light/velocity amplitude ratios and search for a wavelength dependence of
photometric amplitudes which gives us information on limb darkening effects.

The interpretation of these measurements require sophisticated models. Phase
dependent synthetic spectra will be calculated with a code developed by Falter 
(2001). As a prerequisite for these models the time dependent surface 
distribution of $\mathrm{T}_{eff}$, ${\mathrm{log}}\,g$ and the velocity field are
determined using Townsend's (1997) program BRUCE which also takes effects caused
by rotation into consideration. Furthermore, detailed evolutionary pulsation models 
(Charpinet {\it et al.}, these proceedings) are at hand for the comparison with 
the observational frequency pattern.\\

\begin{figure}
\begin{center}
\vspace{8mm}
\epsfig{file=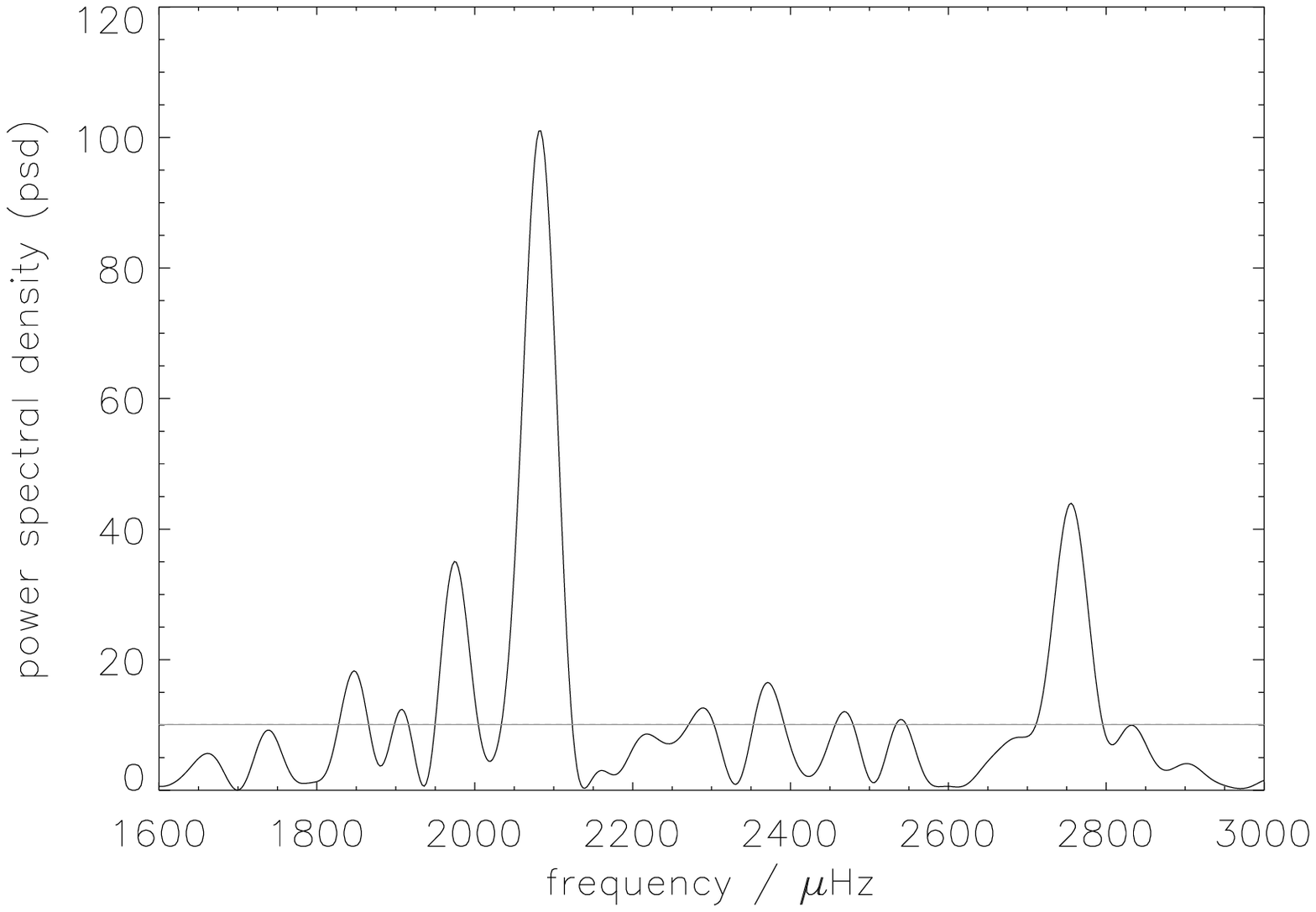,height=35mm,width=35mm,bbllx=200,
bblly=0,bburx=500,bbury=250}
\epsfig{file=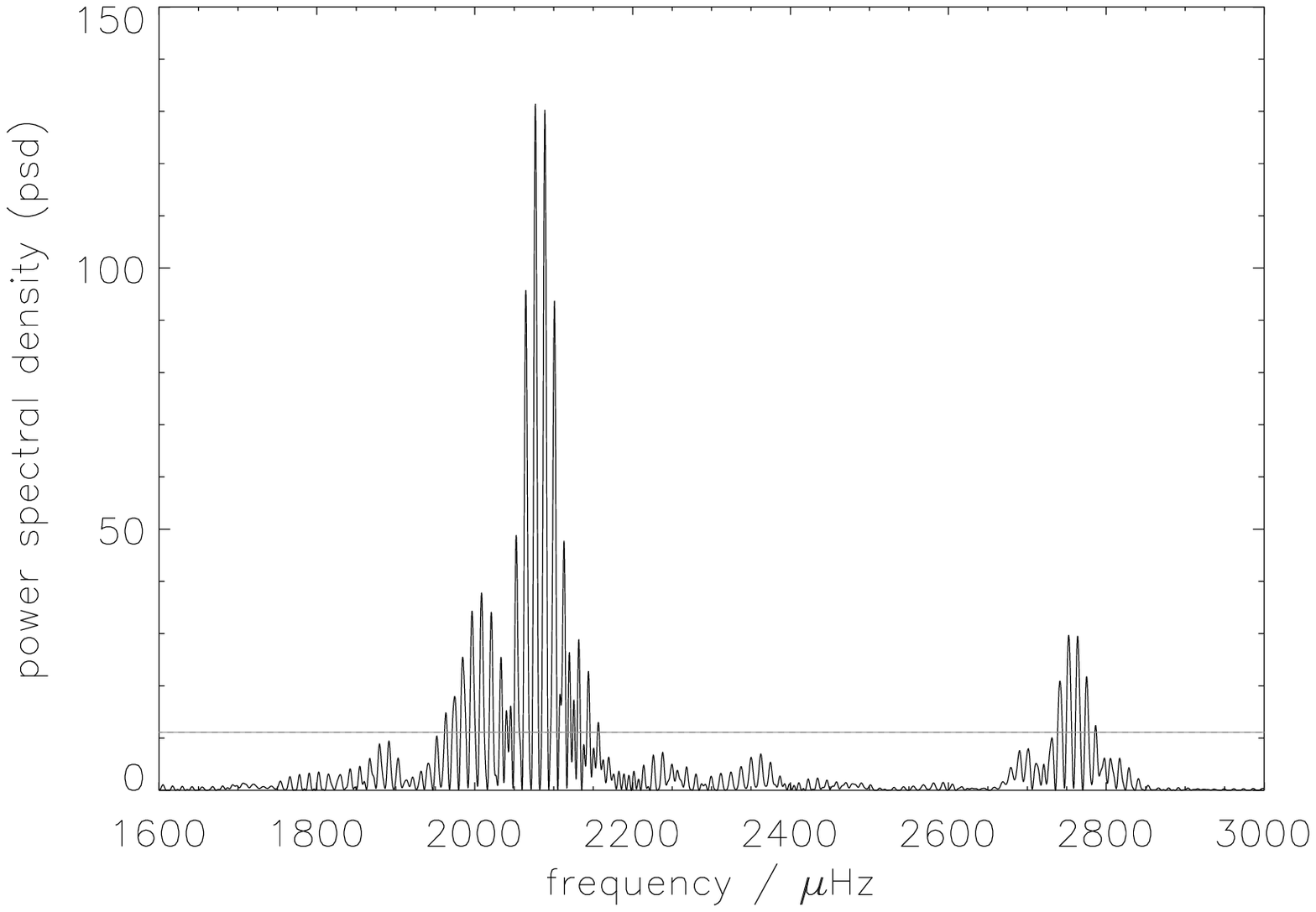,height=35mm,width=35mm,bbllx=0,bblly=0,
bburx=300,bbury=250}
\caption{Periodogram of the RV curve of $\mathrm{H}_{\beta}$ (left panel) and 
of the light curve of the BUSCA 'UV' band (right panel). The power spectral
density (psd) is a measure for the probability that a period is present in the
RV or lightcurve. The horizontal line in both panels is the 3$\sigma$ 
confidence level.}
\label{hbetaf}
\end{center}
\end{figure}

\begin{figure}
\begin{center}
\epsfxsize=7.5cm \epsfbox{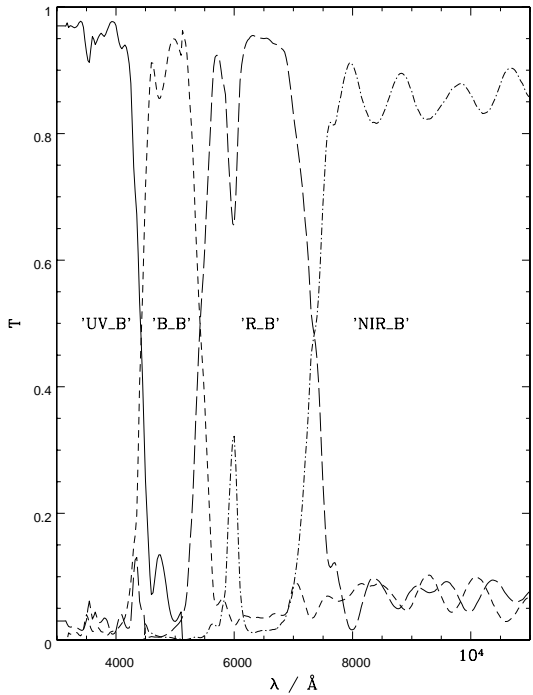}\\[10mm]
\caption{Transmission curves of the four BUSCA wavelength bands. The four ranges
are denoted by 'UV\_B' (central wavelength: 3\,600\,$\mathrm{\AA}$), 'B\_B' 
(4\,800\,$\mathrm{\AA}$), 'R\_B' (6\,300\,$\mathrm{\AA}$) and 'NIR\_B' 
(8\,000\,$\mathrm{\AA}$) which are not comparable to the classical Johnson 
UBVRI system.}
\label{trans}
\end{center}
\end{figure}
\begin{figure}
\begin{center}
\epsfxsize=7.5cm \epsfbox{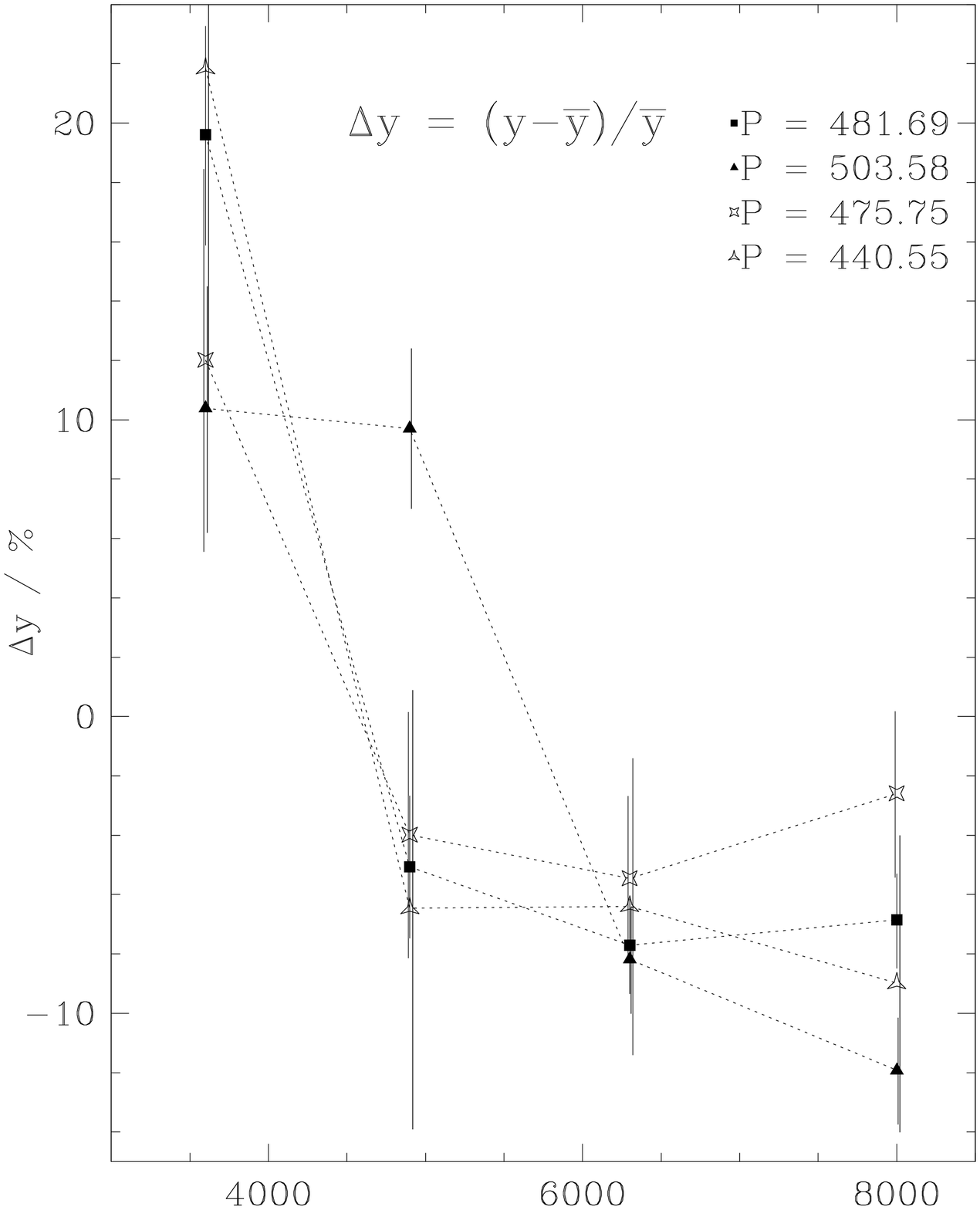}\\[10mm]
\caption{Relative semi amplitudes of four frequencies found with BUSCA. These
are calculated by $\Delta\mathrm{y} = 
\left(\mathrm{y}-\mathrm{\bar{y}}\right)/\mathrm{\bar{y}}$, y denotes the flux
in the wavelength band y and $\mathrm{\bar{y}}$ the mean value of all four
wavelength bands.}
\label{relampl}
\end{center}
\end{figure}
{\bf Feasibility studies:}\\
 As a first step we carried out a feasibility study in May 2001 by monitoring 
simultaneously the radial velocity (RV) curve for about 5.5\,h and the light 
curve in four spectral bands for about 48\,h at Calar Alto Observatory, 
Spain (for details see Falter {\it et al.} 2003).\\
Time resolved optical spectra (spectral resolution: $\sim$1\,$\mathrm{\AA}$)
were obtained at the 3.5\,m telescope equipped with the TWIN spectrograph. The
time resolution was achieved by trailing mode observations which is the tuning 
of the guiding system such that the star moves slowly along the slit in N-S
direction (drift velocity: 270$^{\prime\prime}$, 15 s time resolution). The
periodogram analysis of the RV curves measured for $\mathrm{H}_{\beta}$ and 
$\mathrm{H}_{\gamma}$ reveal three dominant frequencies (see Fig. \ref{hbetaf}). 
These frequencies are
consistent with previous photometry (Kilkenny {\it et al.} 1999). A comparison 
with other feasibility studies (O'Toole {\it et al.} 2000, 2002, Woolf {\it et 
al.} 2002) shows that the power of the modes around 2.1 mHz and 2.75 mHz has 
switched in the course of the years 1999, 2000 and 2001.\\
The light curves were measured with the new multi-band camera BUSCA (see Cordes
{\it et al.} 2003) at the 2.2\,m telescope in four wavelength bands 
simultaneously. Three beamsplitters split the incoming light into four beams 
feeding four 4k$\times$4k CCDs. In order to reduce the achievable S/N and the 
cycle time (51 s) no filters were inserted and, therefore, the beamsplitters 
served as broad wave band filters. Fig. \ref{trans} shows the transmission 
functions of the four bands. The photometric time series in the four wavelength 
bands confirm the three frequencies found in the RV curves and reveal two 
additional frequencies (see Fig. \ref{hbetaf}). These five detections are found 
in all wavebands. More frequencies are present but not consistent in the four 
bands. For all frequencies, we find a wavelength-dependent variation of the
amplitudes. Fig. \ref{relampl} shows the relative deviation of the amplitude in 
the passband from the mean value over all bands (for four example
frequencies).\\

{\bf The Multi-Site Spectroscopic-Telescope project (MSST):}\\
Our feasibility study as well as the previous ones (O'Toole {\it et al.} 2000, 
2002, Woolf {\it et al.} 2002) were unable to resolve the rich power spectrum of
the RV and light curves. A considerably longer period of monitoring with as
little gaps as possible is required to reach this goal.
The next step towards the identification of pulsation modes for PG\,1605$+$072
is the MSST project. From the beginning of May until the end of June in 2002, an 
international collaboration involving more than 25 colleagues measured the light 
and radial velocity curves at 15 observatories around the world. More details 
about that project are described elsewhere (Heber {\it et al.}, these 
proceedings). In the course of this project we additionally had the benefit of a
contribution by the WET collaboration which observed PG\,1605$+$072 as an
alternate target during their Xcov22 campaign (see Schuh {\it et al.} 2003).


\end{document}